\documentclass[aps,amssymb,floatfix,prd,amsmath,preprintnumbers]{revtex4}
\usepackage{amsfonts}
\usepackage{amssymb}
\usepackage{amsmath}
\usepackage{graphicx}  
\usepackage{float}
\usepackage{dcolumn}   
\usepackage{bm}
\begin{document}

\title{\bf  Bouncing cosmology in an Extended Theory of Gravity}

\author{Sunil Kumar Tripathy\footnote{E-mail:tripathy\_ sunil@rediffmail.com}, Rakesh Kumar Khuntia \footnote{E-mail:rakeshkumarkhuntia18@gmail.com}  and Priyabrata Parida \footnote{E-mail:priyabrataparida143@gmail.com}
}
\affiliation{Department of Physics, Indira Gandhi Institute of Technology, Sarang, Dhenkanal, Odisha-759146, India }

\begin{abstract}
We have investigated some bouncing models in the framework of an extended gravity theory where the usual Ricci scalar in the gravitational action is replaced by a sum of the Ricci scalar and a term proportional to the trace of the energy momentum tensor. The dynamical parameters of the model are derived in most general manner. We considered two bouncing scenarios through an exponential and a power law scale factor. The non singular bouncing models also favour a late time cosmic speed up phenomenon. The dynamical behaviour of the equation of state parameter is studied for the models. It is observed that, near the bounce, the dynamics is substantially affected by the coupling parameter of the modified gravity theory as compared to the parameters of the bouncing scale factors.
\end{abstract}
\maketitle
\textbf{PACS number}: 04.50kd.\\
\textbf{Keywords}:  Extended Gravity, Bouncing universe .
\section{Introduction} 

The late time cosmic acceleration of the universe has attracted a lot of research attention in the last two decades after its discovery and further confirmation by a lot of observations from supernova, large scale structure, Baryon Acoustic Oscillation and Cosmic Microwave Background (CMB) radiation anisotropy \cite{Riess98, Perlmutter99, Tegmark2004, Abaz2004, Sperg2003, Hinshaw13, Parkinson12}. The reason behind this late time cosmic dynamics is not yet known exactly. However, researchers have speculated an exotic dark energy(DE) form in the framework of General Relativity (GR) to explain this phenomenon. Observations indicate that DE has a lion share of $68.3\%$ in the mass energy budget of the universe \cite{Ade2016}.  In GR this phenomenon can not be explained in tensor modes and therefore scalar field models are emerged as solution to this intriguing issue. Additional degrees of freedom in the form of scalar fields such as quintessence, phantom fields, ghost fields with unusual Lagrangian are considered in the gravitational action of GR. Chiral cosmological models with multicomponent scalar fields have also been proposed to address this issue \cite{abbya2012, abbya2015, Chervon}. Vector-tensor models with electromagnetic contribution have been proposed which provide some viable solution \cite{Jimenez2008, Jimenez2009, Jimenez2009a}. Amidst all eforts, the very existence of dark energy and its nature still remains as open questions.  Another bizarre fact about the DE is that it can cluster at large scales and violates some energy conditions. While researchers try to get a suitable answer to the late time phenomenon with a good number of DE models, there has also been a debate on whether substantial cosmic acceleration has really occurred in the late epoch or not \cite{Nielsen2016}.

Alternative models to GR have been proposed in recent times as possible solution to the cosmic speed up phenomenon without the need of any DE components and additional dynamical degrees of freedoms as matter fields. In these alternative theories, the geometrical part of the GR field equation is modified in such a manner that, the action will contain some arbitrary function of Ricci Scalar $R$ and possibly the trace of the energy momentum tensor $T$ in place of the Ricci scalar appearing in GR. The geometrically modified gravity theories have been motivated from quantum effects and are able to provide the necessary acceleration without requiring any additional dynamical fields like quintessence or ghost fields. Several modified theories of gravity have been proposed in literature such as $f(R)$ theory \cite{Caroll2004, Nojiri2007}, $f(G)$ gravity \cite{Nojiri2005}, $f(\mathcal{T})$ theory \cite{Linder2010, Myrza2011} and $f(R,T)$ theory \cite{Harko2011}.  There have been a wide interest and investigations concerning issues in cosmology and astrophysics using the recently proposed $f(R,T)$ gravity theory \cite{Mishra16, Mishra18c, PKS2018, Sahu2017, Yousaf16, Tretyakov18, Velten17, Carvalho17,  Baffou19, Alha16, Alves16,Abbas17}.   As a simple extension of the $f(R,T)$ theory, extended theories of gravity have been proposed which provide a simple structure but have elegance in addressing  many issues in cosmology \cite{ Capo2011,Mishra18a, Mishra18b, Capo2019}. 

The standard cosmological model provides a good explanation of the early universe but suffers from issues like flatness problem, the horizon problem, initial singularity and baryon asymmetry problem. The inflationary scenario solved some of these issues of the early universe standard model and provided a causal theory of structure formation \cite{Guth1981, Mukhanov1981}. However, the inflationary scenario suffers from the singularity problem and the trans-Planckian problem for fluctuations. Before the onset of inflation where the universe undergoes an almost exponential expansion, singularity occurs and therefore inflationary scenario fails to reconstruct the complete past history of the universe. Matter bounce scenario have been proposed as a solution to the challenges faced by the inflationary scenario \cite{Brand2011, Bars2011, Bars2012}. For some reviews on bouncing cosmologies one may refer to \cite{Brand2012, Bate2014, Novello2008, Brand16}. In bouncing scenario, the universe undergoes an initial matter dominated contraction phase followed by a non-singular bounce and then there is a causal generation for fluctuation. Bouncing cosmologies have been investigated in alternative gravity theories such as $f(R)$ theory \cite{Bamba2014, Barag2009, Barag2010, Saidov2010, Saikat2018}, modified Gauss-Bonnet gravity \cite{Bamba2014a, Bamba2015}, $f(R,T)$ gravity \cite{Singh18} and $f(T)$ gravity \cite {Cai2011}.

The present work reports the investigation of  some bouncing models in the framework of an extended theory of gravity. For this purpose we have considered a simple extended gravity theory with an isotropic FRW universe.  The organisation of the paper is as follows: Section II contains the basic formalism of extended gravity where we have derived the field equations and expressed the dynamical parameters of the model in a general manner in terms of the Hubble rate. In Section III, two bouncing models are investigated by assuming an exponential and a power law scale factors showing bouncing behaviour. The effect of the model parameters on the cosmic dynamics have been discussed.  Section-IV contains a brief summary and conclusion of the present work.

\section{Basic Formalism}
We consider the action for a geometrically modified theory of gravity as 
\begin{equation} \label{eq:1}
S=\int d^4x\sqrt{-g}\left[\frac{1}{2} f(R,T)+ \mathcal{L}_m \right],
\end{equation}
where $f(R,T)$ is an arbitrary function of the Ricci scalar $R$ and the trace $T$ of the energy-momentum tensor, $\mathcal{L}_m$ is the usual matter Lagrangian.  The action reduces to that of GR for $f(R,T)=R$.  We use the natural system of unit: $8\pi G=c=1$; $G$ and $c$ are respectively the Newtonian gravitational constant and speed of light in vacuum.

For minimal matter-geometry coupling, we can have $f(R,T)=f_1(R)+f_2(T)$ so that the action becomes
\begin{equation} \label{eq:2}
S=\int d^4x\sqrt{-g}\left[\frac{1}{2} \left(f_1(R)+f_2(T)\right)+ \mathcal{L}_m \right].
\end{equation}

If we vary the action with respect to the metric $g_{\mu\nu}$, the field equations are obtained for the modified theory of gravity as

\begin{equation} \label{eq:3}
R_{\mu\nu}-\frac{1}{2}f^{-1}_{1,R} (R)f_1(R)g_{\mu\nu}=f^{-1}_{1,R}(R)\left[\left(\nabla_{\mu} \nabla_{\nu}-g_{\mu\nu}\Box\right)f_{1,R}(R)+\left[1 +f_{2,T}(T)\right]T_{\mu\nu}+\left[f_{2,T}(T)p+\frac{1}{2}f_2(T)\right]g_{\mu\nu}\right].
\end{equation}

Here the matter Lagrangian is assumed to be proportional to the pressure $p$ of the cosmic fluid i.e. $\mathcal{L}_m=-p$. Also, we have adopted the shorthand notations
\begin{equation}\label{eq:4}
f_{1,R} (R)\equiv \frac{\partial f_1(R)}{\partial R},~~~~~~~~~ f_{2,T} (T)\equiv \frac{\partial f_2(T)}{\partial T}, ~~~~~~~~~ f^{-1}_{1,R} (R) \equiv \frac{1}{f_{1,R} (R)}.
\end{equation}
The energy-momentum tensor $T_{\mu\nu}$ is related to the matter Lagrangian as
\begin{equation}\label{eq:5}
T_{\mu\nu}=-\frac{2}{\sqrt{-g}}\frac{\delta\left(\sqrt{-g}\mathcal{L}_m\right)}{\delta g^{\mu\nu}}.
\end{equation}

Considering a simple choice $f_1(R)=R$, we obtain
\begin{equation}\label{eq:6}
G_{\mu\nu}= \left[1 +f_{2,T}(T)\right]T_{\mu\nu}+\left[f_{2,T}(T)p+\frac{1}{2}f_2(T)\right]g_{\mu\nu},
\end{equation}
which can also be written as
\begin{equation}\label{eq:7}
G_{\mu\nu}= \kappa_{T}\left[T_{\mu\nu}+ T^{int}_{\mu\nu}\right].
\end{equation}
Here, $G_{\mu\nu}= R_{\mu\nu}-\frac{1}{2}Rg_{\mu\nu}$ is the usual Einstein tensor and $\kappa_{T}= 1 +f_{2,T}(T)$ is the redefined Einstein constant.  In \eqref{eq:7}, we have the effective energy-momentum tensor generated due to the geometrical modification through a minimal coupling with matter,

\begin{equation}\label{eq:8}
T^{int}_{\mu\nu}=\left[ \frac{f_{2,T}(T)p+\frac{1}{2}f_2(T)}{1 +f_{2,T}(T)}\right]g_{\mu\nu},
\end{equation}
A suitable choice of $f_2(T)$ may provide a viable cosmological model which may be confronted with recent observations.

In the present work, we consider a linear functional 

\begin{equation}\label{eq:9}
\frac{1}{2}f_2(T)=\lambda T,
\end{equation}
so that $\kappa_T = 1+2\lambda$ and $T^{int}_{\mu\nu} = \frac{g_{\mu\nu}}{\kappa_T}\left[\left(2p+T\right)\lambda\right]$.

We consider a flat FRW  metric
\begin{equation}
ds^2 = dt^2 - a^2(t)(dx^2+dy^2+dz^2),
\end{equation}
for the investigation of some bouncing models. The Friedman equations for a perfect fluid distribution in the universe with $T_{\mu\nu}=-pg_{\mu\nu}+\left(\rho+p\right)u_{\mu}u_{\nu}$ are obtained as,
\begin{equation}
3H^{2} = (1+3\lambda)\rho - \lambda p,
\end{equation}
\begin{equation}
2\dot{H} + 3H^{2} = \lambda \rho - (1+ 3\lambda)p.
\end{equation}
Here  $a = a(t)$ is the scale factor of the universe and $ H = \frac{\dot{a}}{a}$ is the Hubble parameter. The overhead dot denotes differentiation with respect to $t$.

From the above field equations we can determine the pressure, energy density and equation of state (EoS) parameter in terms of the Hubble parameter. The pressure and the energy density for the present model are obtained as,
\begin{eqnarray}
p &=& -\frac{ 3(1 + 2\lambda)H^{2} + 2 (1 + 3\lambda) \dot{H}}{(1 + 3\lambda)^{2} - \lambda^{2}},\\
\rho &=&  \frac{3(1 + 2\lambda)H^{2} - 2 \lambda \dot{H}}{(1 + 3\lambda)^{2} - \lambda^{2}}.
\end{eqnarray}

The EoS parameter $\omega=\frac{p}{\rho}$ becomes,
\begin{equation}
\omega =  -\frac{3(1 + 2\lambda)H^{2}+ 2 (1 + 3\lambda) \dot{H}}{3(1 + 2\lambda)H^{2} - 2\lambda \dot{H}}.
\end{equation}

It is obvious from the above expressions that, the evolutionary behaviour of the dynamical parameters such as the pressure, energy density and the EoS parameter depend on the model parameter $\lambda$ and the evolutionary aspect of the Hubble parameter. One can note that, for a vanishing $\lambda (\lambda=0)$, the EoS parameter reduces to that in GR: $\omega=-1-\frac{2}{3}\frac{\dot{H}}{H^2}$.

\section{Bouncing Cosmologies}
In the present work, we are interested to investigate some bouncing models in the framework of an extended theory of gravity and to study the cosmic dynamics through the dynamical properties such as energy density, pressure and equation of state parameter. We have obtained the expressions for these quantities in terms of the Hubble parameter. Here we have considered two different bouncing models described by certain scale factors and studied the consequent dynamical behaviour of the models.

\subsection{Model-I}
A symmetric bounce can be modelled through the scale factor,
\begin{equation}
a = e^{\beta t^{2}}\label{eq:4.1}
\end{equation}
where, $\beta$ is a positive constant parameter that controls the cosmic expansion. The Hubble's parameter  for this scale factor is obtained as,
\begin{equation}
H = 2\beta t.
\end{equation}\
The bouncing scenario occurs at $t=0$ and obviously at bounce, we have  $H=0$ and $\dot{H} = 2\beta >0$.

The bouncing scale factor is shown as a function of cosmic time for three representative values of $\beta$ in Figure 1. It is observed from the figure that, scale factor is symmetric about the bouncing epoch $t=0$. The slope of the curves of the scale factor is proportional to the parameter $\beta$. In other words, more is the value of $\beta$, more is the curvature of the plot. In Figure 2, the time evolution of the Hubble parameter is shown for three representative values of the parameter $\beta$. The Hubble parameter is a linear function of cosmic time. Since we are dealing with a bouncing model, the cosmic time may range from negative domain to positive domain and consequently the Hubble parameter evolves linearly from negative time domain to positive domain through $H=0$ at bounce. The slopes of the straight lines in Fig 2 are proportional to $\beta$. 

\begin{figure}[t]
\centering
\includegraphics[width=0.7\textwidth]{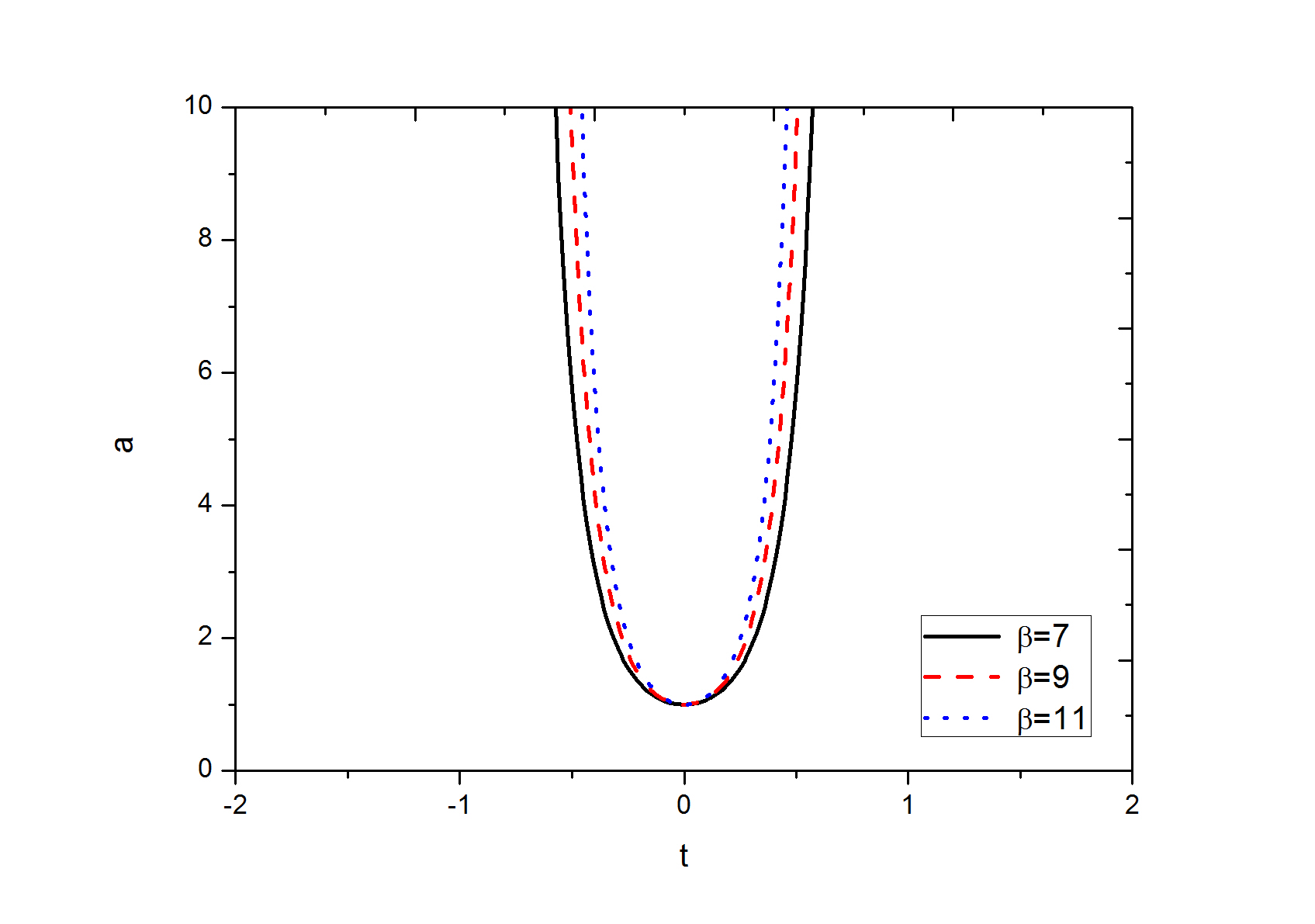}
\caption{Evolutionary behaviour of the scale factor for three representative values of $\beta$ for Model-I.}
\label{fig:scaleavst}
\end{figure}

\begin{figure}[htbp]
\centering
\includegraphics[width=0.7\textwidth]{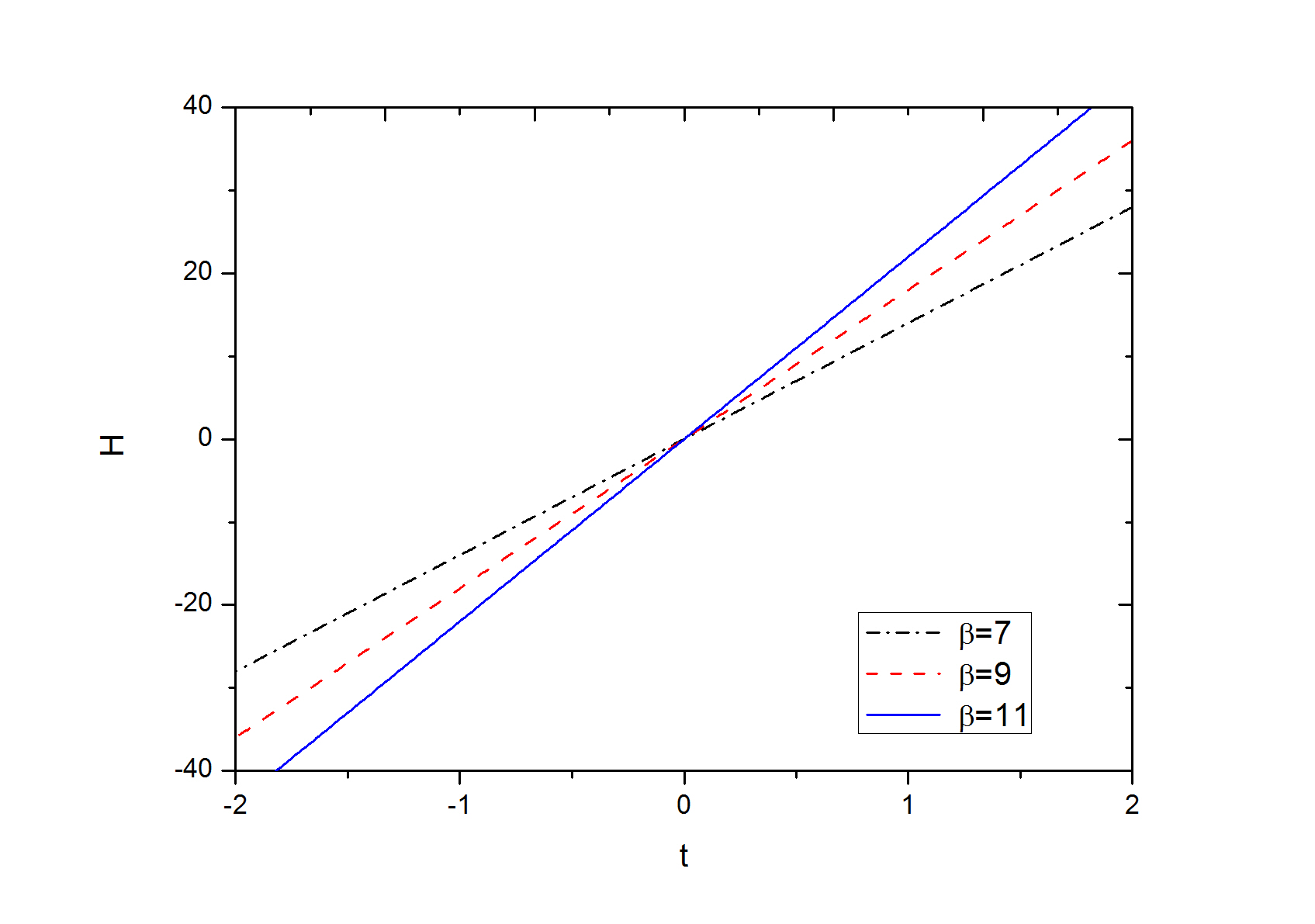}
\caption{Evolutionary behaviour of Hubble parameter for three representative values of $\beta$ for Model-I.}
\label{fig:scalehvst}
\end{figure}

\begin{figure}[htbp] 
\centering
\includegraphics[width=0.7\textwidth]{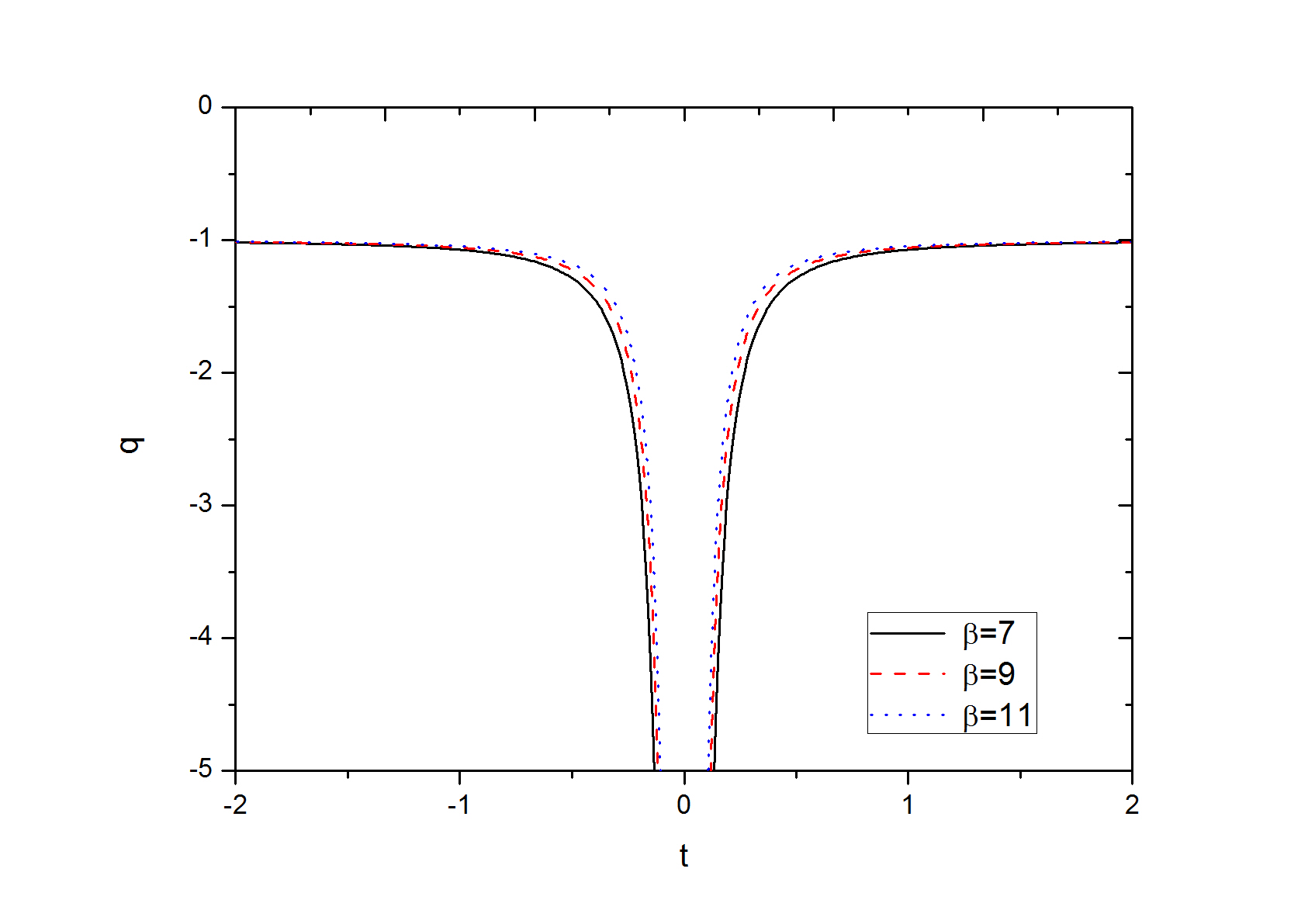}
\caption{Evolutionary behaviour of deceleration parameter for three representative values of $\beta$ for Model-I.}
\label{fig:scaleqvst}
\end{figure}

The deceleration parameter(DP) $q$ is an important quantity in the study of cosmic dynamics. It is defined as
\begin{equation}
q=-1-\frac{\dot{H}}{H^2},
\end{equation}
which for the present bouncing model becomes
\begin{equation}
q=-1-\frac{1}{2\beta t^2}.\label{eq:19}
\end{equation}
A positive value of $q$ signifies a decelerated universe and a negative value of it explains an accelerated universe. One can note from the above eq. \eqref{eq:19} that the deceleration parameter remains negative for all values of cosmic time. In other words, the present bouncing model always predicts an accelerated universe, which may be in conformity with observations  at least at the present epoch.  The deceleration parameter is independent of the choice of the parameter $ \lambda $ and depends only on $ \beta $. In Figure 3, the time evolution of the deceleration  parameter is shown for three representative values of the parameter $\beta$. It is observed from the figure that, deceleration parameter is symmetric about the bouncing time $t=0$. In the negative time zone, DP evolves from $q=-1$ to large negative values near the bounce. In the positive time zone, it evolves from large negative values to $q=-1$ at late times.

The energy density $\rho$ and pressure $p$ for the present model are obtained as

\begin{eqnarray}
\rho &=& \frac{12\beta^{2}(1 + 2 \lambda)t^{2} - 4\beta \lambda }{(1 + 3\lambda)^{2} - \lambda^{2}},\\
p &=& -\frac{12\beta^{2}(1 + 2 \lambda)t^{2} + 4\beta (1 + 3\lambda)}{(1 + 3\lambda)^{2} - \lambda^{2}}.
\end{eqnarray}

It is clear from the above expressions that, the evolutionary behaviour of the energy density and pressure depends on the value of the scale factor parameter $\beta$ and  model parameter $\lambda$. Since the denominators of $\rho$ and $p$ are positive for a given $\lambda >0$, the positivity or the negativity of these quantities depend only on the respective numerators. In order to satisfy certain energy conditions, the energy density should remain positive throughout the cosmic evolution. This can be achieved only when $\beta t^2 > 0.33 (\frac{1}{\lambda}+2)^{-1}$. At late phase of cosmic evolution, this condition is easily satisfied. However, at a cosmic epoch around the bounce, there is uncertainty whether this condition will be satisfied or not. In view of this, in the present work, we have considered suitable values of $\beta$ so that the energy density remains positive through out the cosmic evolution both in the positive and negative domains of cosmic time. From a  systematic investigation, we have found three different values of $ \beta $ i.e. $ \beta =  7,~ 9$ and 11 and three different value of $\lambda $ such as $0$, $0.01$ and 0.1, for which we get positive energy density through out the cosmic evolution.  In choosing these values of $\beta$ and $\lambda$, we are not ruling out any other possible values of $\beta$ and $\lambda$ that can satisfy the above positive energy condition.

The EoS parameter can be obtained from the expressions of pressure and energy density as
\begin{equation}
\omega = -\frac{12\beta^{2}(1 + 2 \lambda)t^{2} + 4\beta (1 + 3\lambda)}{12\beta^{2}(1 + 2 \lambda)t^{2} - 4\beta \lambda} .
\end{equation}
Since energy density is constrained to remain positive at all epochs and the pressure is negative throughout the cosmic evolution, it is obvious that the EoS parameter becomes a negative quantity. The evolutionary behaviour of the equation of state parameter can be assessed through the specific choices the model parameters $\lambda$ and $\beta$. We have shown the dynamical evolution of $\omega$ for the constrained values of $\beta$ and $\lambda$ in Figures 4-6. In Figure 4, the  EoS parameter is shown for three representative values of the parameter $\beta$ for $\lambda=0$. It is needless to mention that, model corresponding to $\lambda=0$ reproduces the results of GR i.e. $\omega_{\lambda\rightarrow 0} \simeq -1+\frac{1}{3\beta t^2}$.  $\omega$ for this case is symmetric about the bouncing epoch and evolves in the phantom region from a large negative value near the bounce to an asymptotic value of $-1$. The EoS parameter evolves rapidly near the bounce as compared to the time frame far away from bounce. One can observe from the figure that, the scale factor parameter $\beta$ has a role to raise the rate of increment in $\omega$ near the bounce. It is interesting to note that, $\beta$ does not have an impressive impact on $\omega$ at an epoch far away from the bounce.
\begin{figure}[htbp] 
\centering
\includegraphics[width=0.7\textwidth]{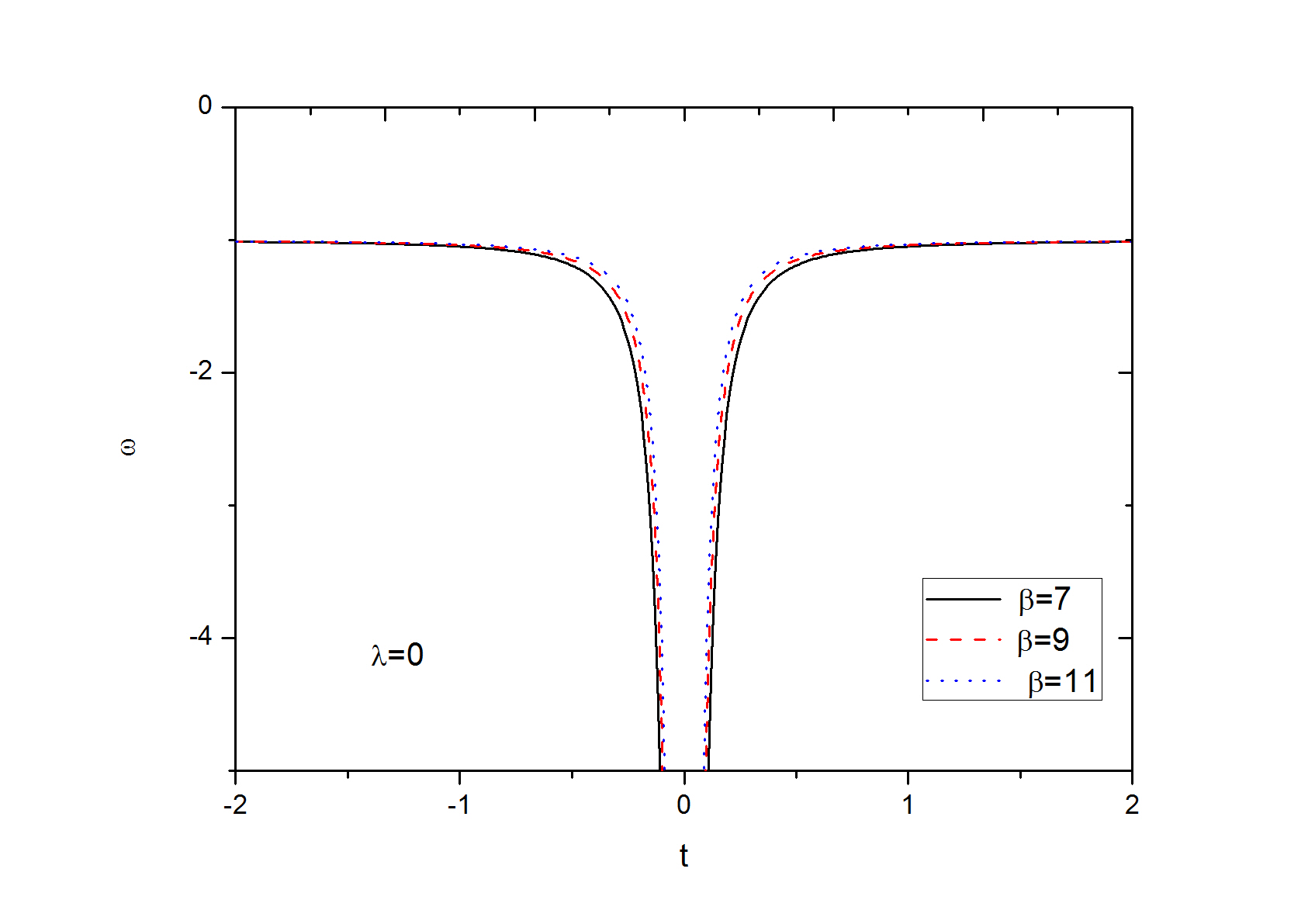}
\caption{Evolutionary behaviour of equation of state parameter for three representative values of $\beta$ for Model-I.}
\end{figure}
\begin{figure}[htbp] 
\centering
\includegraphics[width=0.7\textwidth]{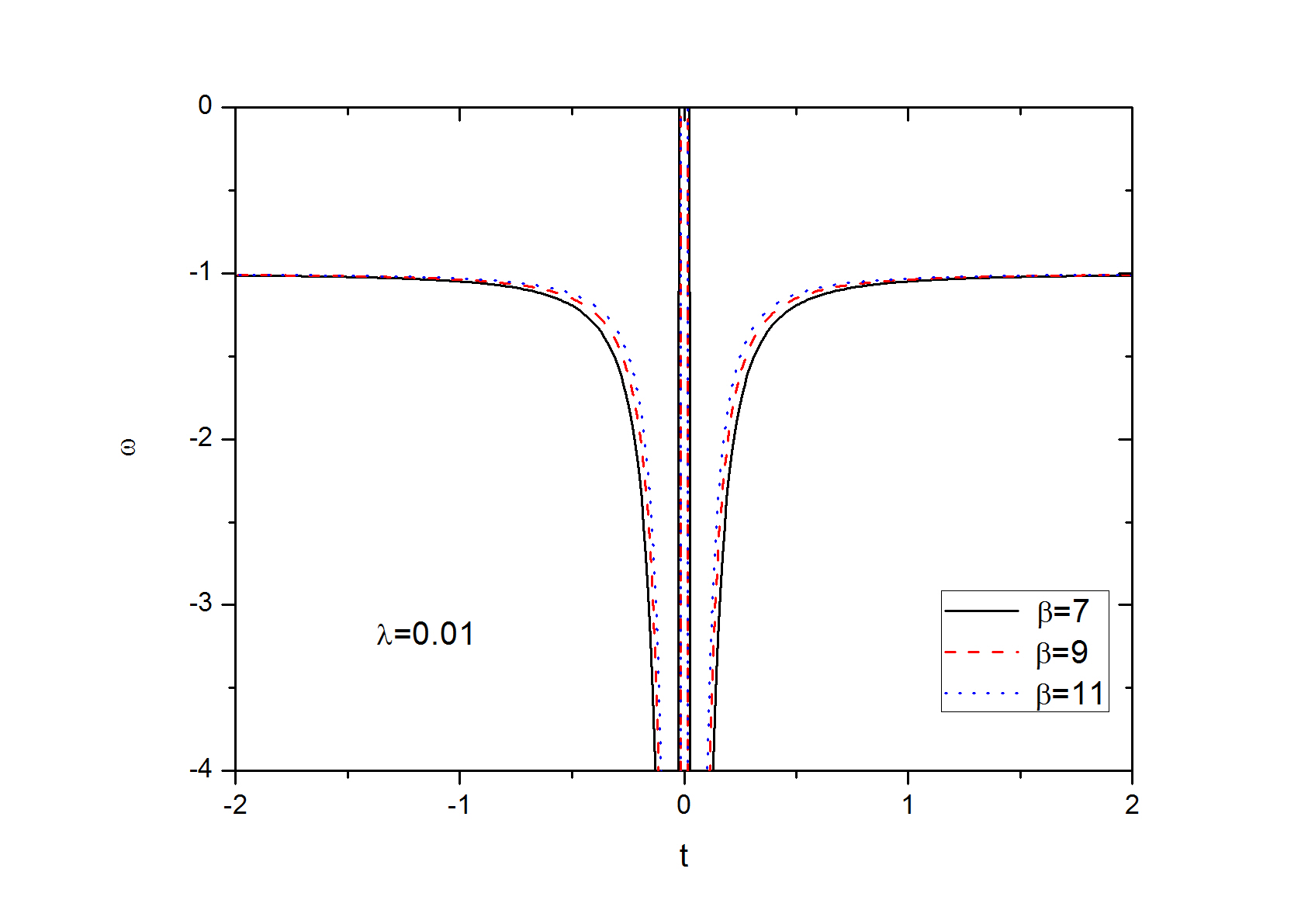}
\caption{Evolutionary behaviour of equation of state parameter for three representative values of $\beta$ for Model-I.}
\end{figure}
\begin{figure}[htbp] 
\centering
\includegraphics[width=0.7\textwidth]{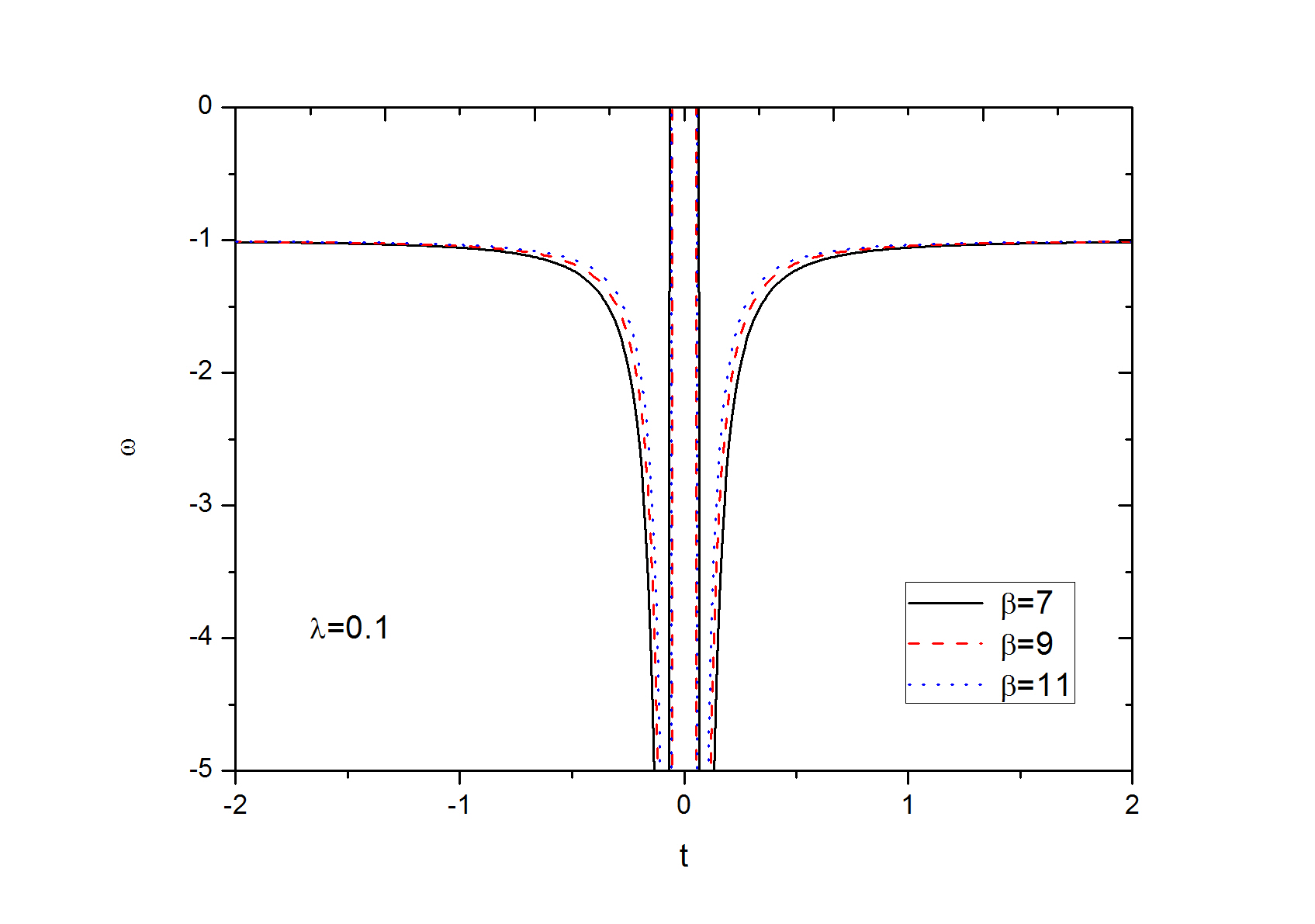}
\caption{Evolutionary behaviour of equation of state parameter for three representative values of $\beta$ for Model-I.}
\end{figure}
\begin{figure}[htbp] 
\centering
\includegraphics[width=0.7\textwidth]{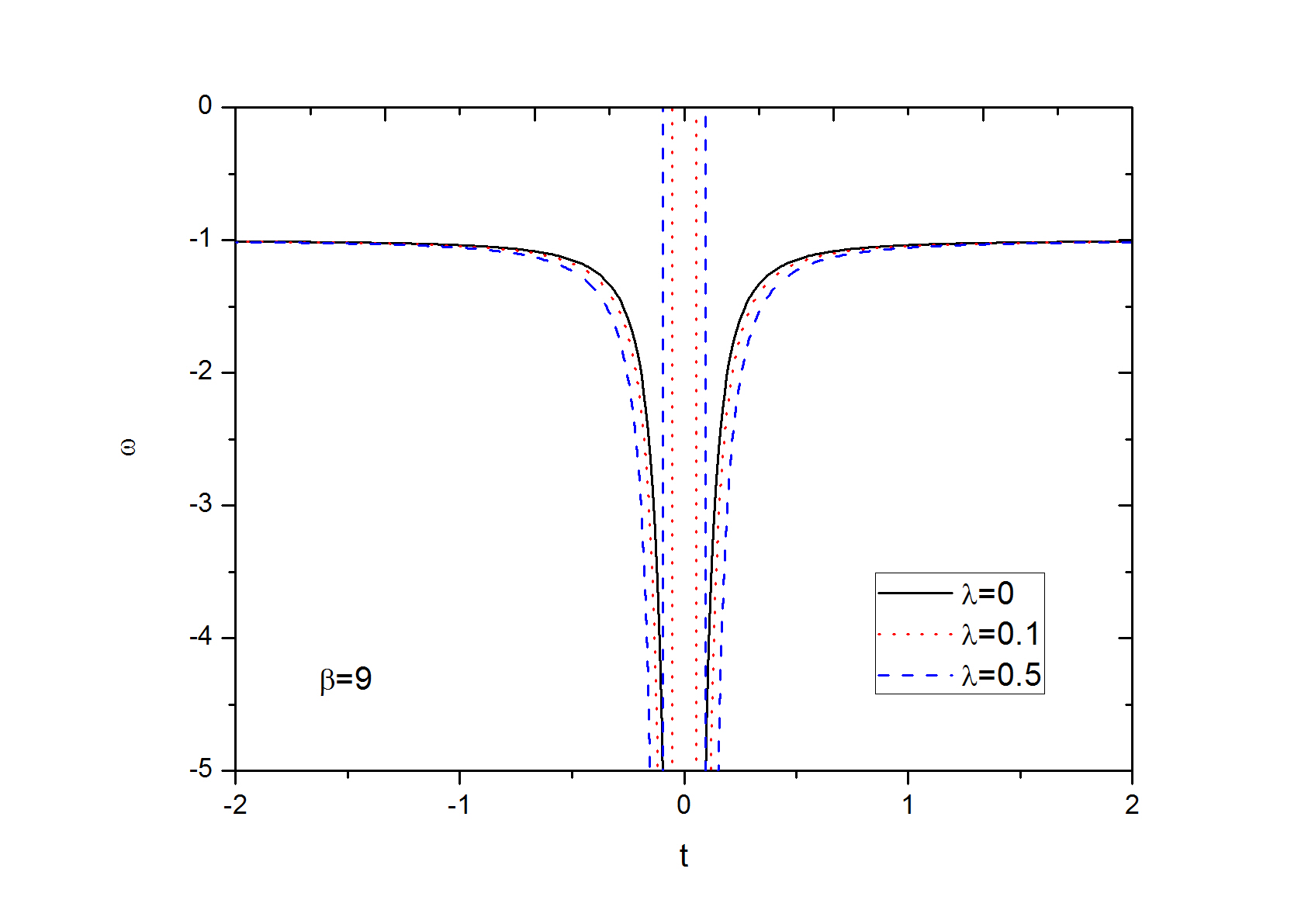}
\caption{Evolutionary behaviour of equation of state parameter for three representative values of $\lambda $ for Model-I.}
\end{figure}

In Figures 5 and 6, we have considered some finite values of $\lambda$ namely $0.01$ and $0.1$. Since $\lambda \neq 0$, we can have an analytic expression for the EoS parameter at bounce. In fact, at the bouncing epoch, the EoS parameter becomes $\omega_{B} = \frac{1}{\lambda}+3$ which is a positive quantity independent of the choice of $\beta$. This is reflected in the figures where we have some vertical lines near bounce region.

In order to assess the effect of the model parameter $\lambda$ on the dynamical aspects of the EoS parameter, in Figure 7, we have plotted $\omega$ for different values of $\lambda$ for a given value of $\beta$. As is evident from the figure, $\lambda$ has an appreciable effect on $\omega$ near the bouncing epoch than at a far away time frame. Near the bouncing epoch, the role of $\lambda$ is to either increase or decrease the rate of increment in $\omega$. In other words, lower the value of $\lambda$, the lower the time $\omega$ takes to reach the asymptotic value of $-1$.

It is certain that, the present model evolves in the phantom region and therefore some of the energy conditions are violated. The null energy condition for the present model can be expressed as
\begin{equation}
\rho+p=-\frac{4\beta(1+4\lambda)}{(1 + 3\lambda)^{2} - \lambda^{2}}.
\end{equation}
Since we have chosen the parameters $\beta$ and $\lambda$ to be positive, we have $\rho+p <0$ which signifies a violation of the null energy condition $\rho+p \geq 0$.
\subsection{Model-II}
We consider another  bouncing scale factor \cite{Molina1999},
\begin{equation}
a(t) = \sqrt{a_{0}^{2} + \beta^{2}t^{2}},
\end{equation}
where, $\beta$ is a positive constant parameter and $a_{0}$ is the radius scale factor at bounce. With proper renormalization of the constants $a_0$ and $\beta$, one may get some quintom behaviour of the model.  In this model bounce occurs at $t=0$. We have plotted the bouncing scale factor for three representative values of $\beta$ namely 0.9, 1 and 1.12 in Figure 8 where  $a_{0}=1$ is assumed. The scale factor is symmetric about $t=0$. The parameter $\beta$ controls the slope of the scale factor. Higher the value of $\beta$, higher is the slope.  


\begin{figure}[h!]
\centering
\includegraphics[width=0.7\textwidth]{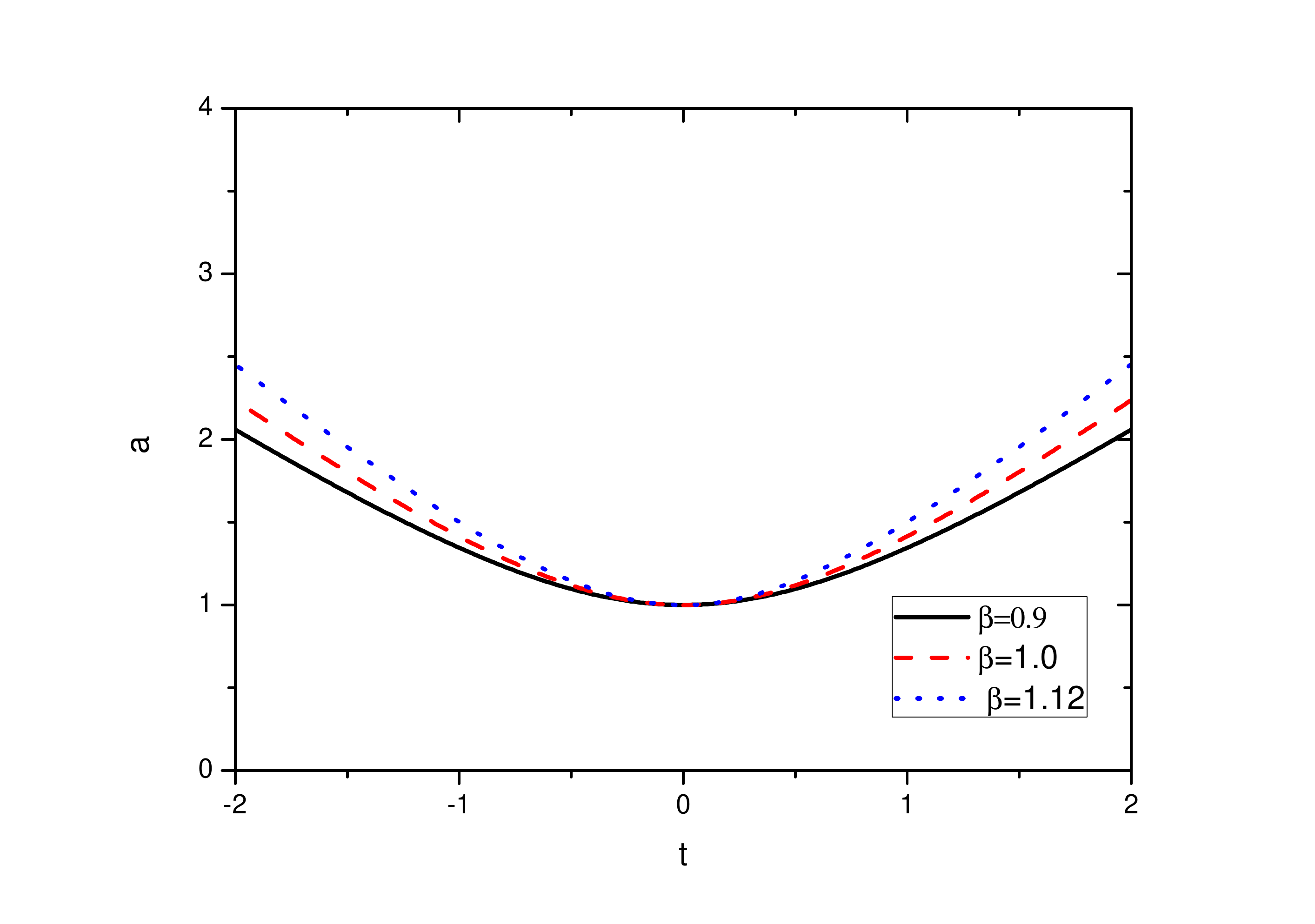}
\caption{Scale factor for three representative values of $\beta$ for Model -II.}
\end{figure}

For the above scale factor, the Hubble's parameter is obtained as,
\begin{eqnarray}
H =\frac{\beta ^{2}t}{ 1+\beta ^{2}t^{2}},
\end{eqnarray}
and its first derivative as
\begin{equation}
\dot{H}=\beta^{2}\left[\frac{1 - \beta^{2}t^{2}}{\left(1 + \beta^{2}t^{2}\right)^{2}}\right].
\end{equation}

\begin{figure}[h!]
\centering
\includegraphics[width=0.7\textwidth]{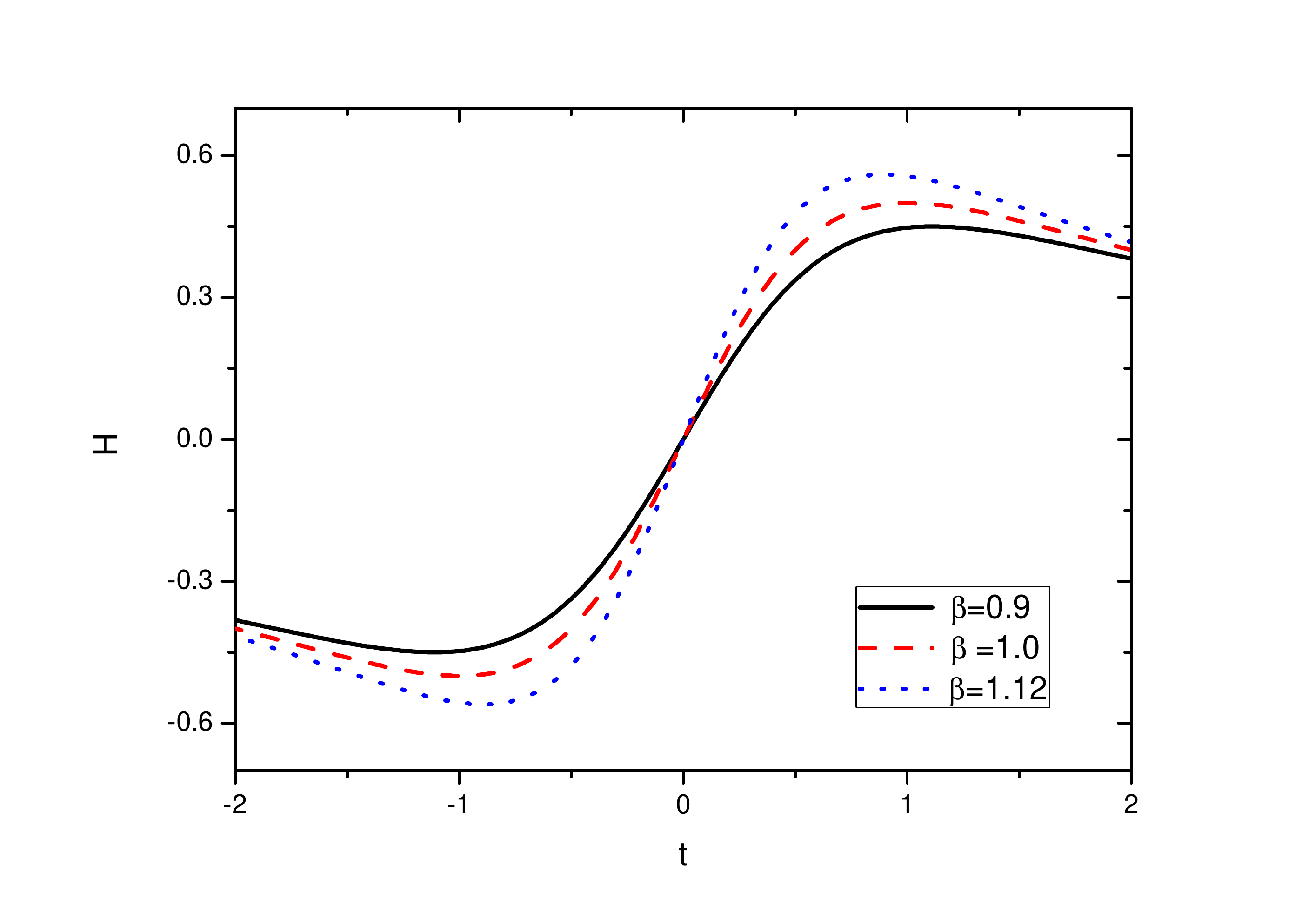}
\caption{Evolutionary behaviour of Hubble's parameter for three representative values of $\beta$ for Model -II.}
\end{figure}
In Figure 9, the Hubble parameter is shown for three representative values of $\beta$. The DP for the present bouncing model is given by
\begin{eqnarray}
q = -\left(\frac{1}{\beta t}\right)^{2}.
\end{eqnarray}
In Figure 10, the evolutionary behaviour of DP is shown for three representative values of $\beta$.  For all positive values of $\beta$ it becomes negative which signifies an accelerated universe. 

\begin{figure}[h]
\centering
\includegraphics[width=0.7\textwidth]{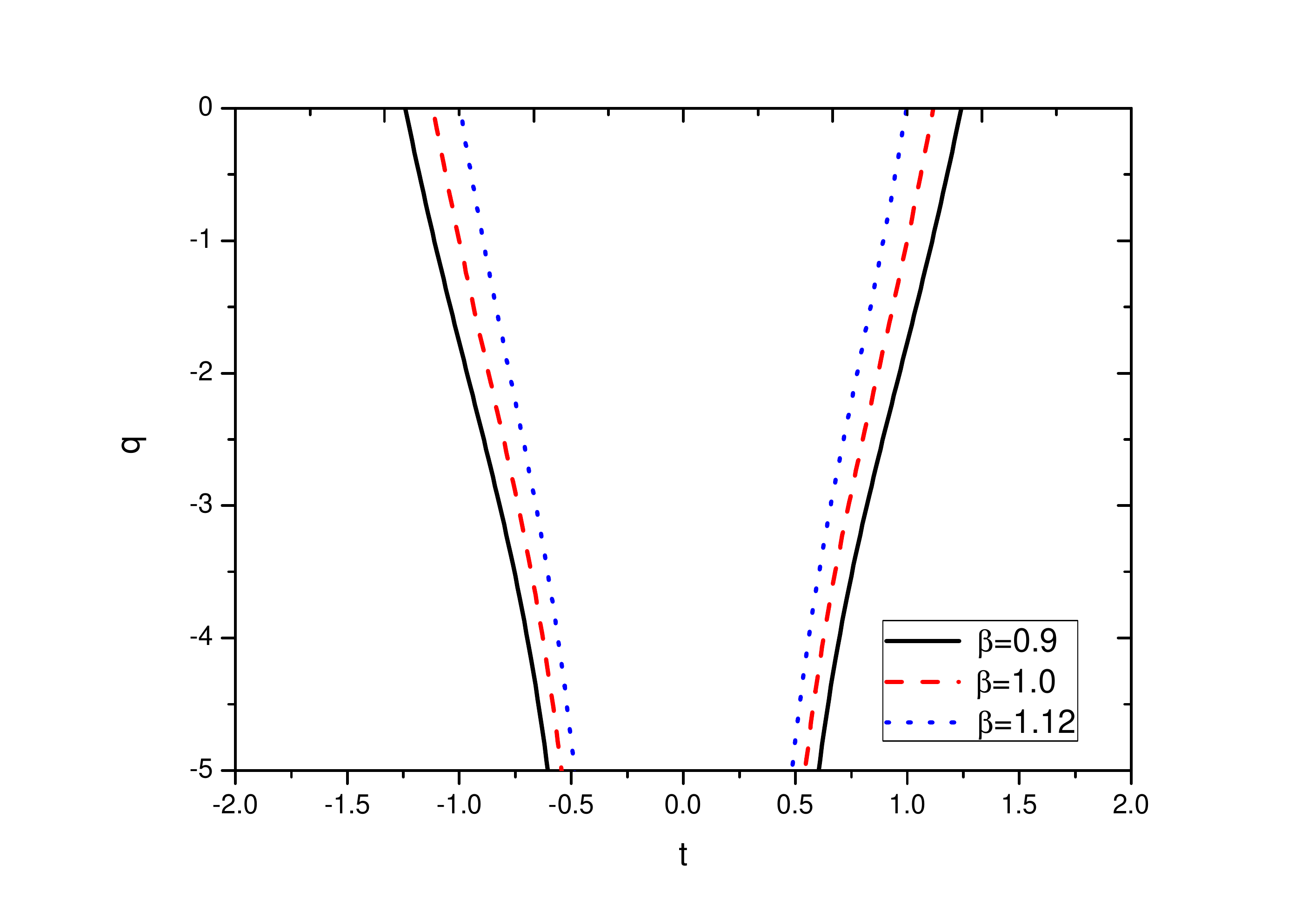}
\caption{Evolutionary behaviour of deceleration  parameter for three representative values of $\beta$ for Model -II.}
\end{figure}

The energy density $\rho$ and pressure $p$ for the present model are obtained as
\begin{eqnarray}
\rho &=& \frac{(3+8\lambda)\beta^4t^2-2\lambda \beta^{2}}{(1 + \beta^{2}t^{2})^{2}[{( 1 + 3\lambda)^{2} - \lambda^{2}}]},\\
 p &=& -\frac{\beta^{4}t^{2}+2(1+3\lambda)\beta^2}{
   (1 + \beta^{2}t^{2})^{2}[{( 1 + 3\lambda)^{2} - \lambda^{2}}]}.
\end{eqnarray}

It is clear from the above expressions that, in this model also the evolutionary behaviour of the energy density and pressure depend on the value of the scale factor parameter $\beta$ and  model parameter $\lambda$. The denominators of both the expression are always positive for a positive value of $\lambda$. The positivity or the negativity of these quantities depend only on the respective numerators. In order to satisfy certain energy conditions, the energy density should remain positive throughout the cosmic evolution. At a cosmic epoch around the bounce, there is uncertainty whether this condition will be satisfied or not. In view of this, in the present work, we have considered suitable values of $\beta$ so that the energy density remains positive through out the cosmic evolution both in the positive and negative domains of cosmic time. We consider three different values of scale parameter $ \beta $ i.e. $ \beta $=0.9, 1 and 1.12 and three different value of model parameter $\lambda $ namely 0, 0.1 and  0.5.

The equation of state parameter $ \omega $  for the present case becomes
\begin{eqnarray}
\omega  &=& -\frac{\beta^{4}t^{2}+2(1+3\lambda)\beta^2}{(3+8\lambda)\beta^4t^2-2\lambda \beta^{2}}.
\end{eqnarray}

The equation of state parameter is a negative quantity and depends on the model parameters $\beta$ and $\lambda$. In Figures 11-14,  the dynamical  behaviour of the equation of state parameter is shown for the constrained values of the parameters $\beta$ and $\lambda$. In Fig 11, we have considered $\lambda=0$ which reproduces the behaviour in GR. $\omega$ evolves from large negative value near the bounce to rapidly becoming positive as it moves away from bouncing epoch. The rate of increment of $\omega$ increases as we increase the value of $\beta$ for a given non zero finite value of $\lambda$. However, one can observe from Fig 14 that, for a given value of $\beta$, the rate of increment of $\omega$ decreases with the increase in $\lambda$. The effect of $\beta$  on the equation of state parameter is quite visible at a time frame away from bounce than near the bouncing epoch. For finite value of $\lambda$, the equation of state parameter assumes a value of $\omega_{B} \simeq \frac{1}{\lambda}+3$ which is the same we have obtained in  model-I. It is interesting to note that in this case also, the EoS parameter at bounce does not depend on the scale factor parameter $\beta$. 


\begin{figure}
\minipage{0.40\textwidth}
\centering
\includegraphics[width=65mm]{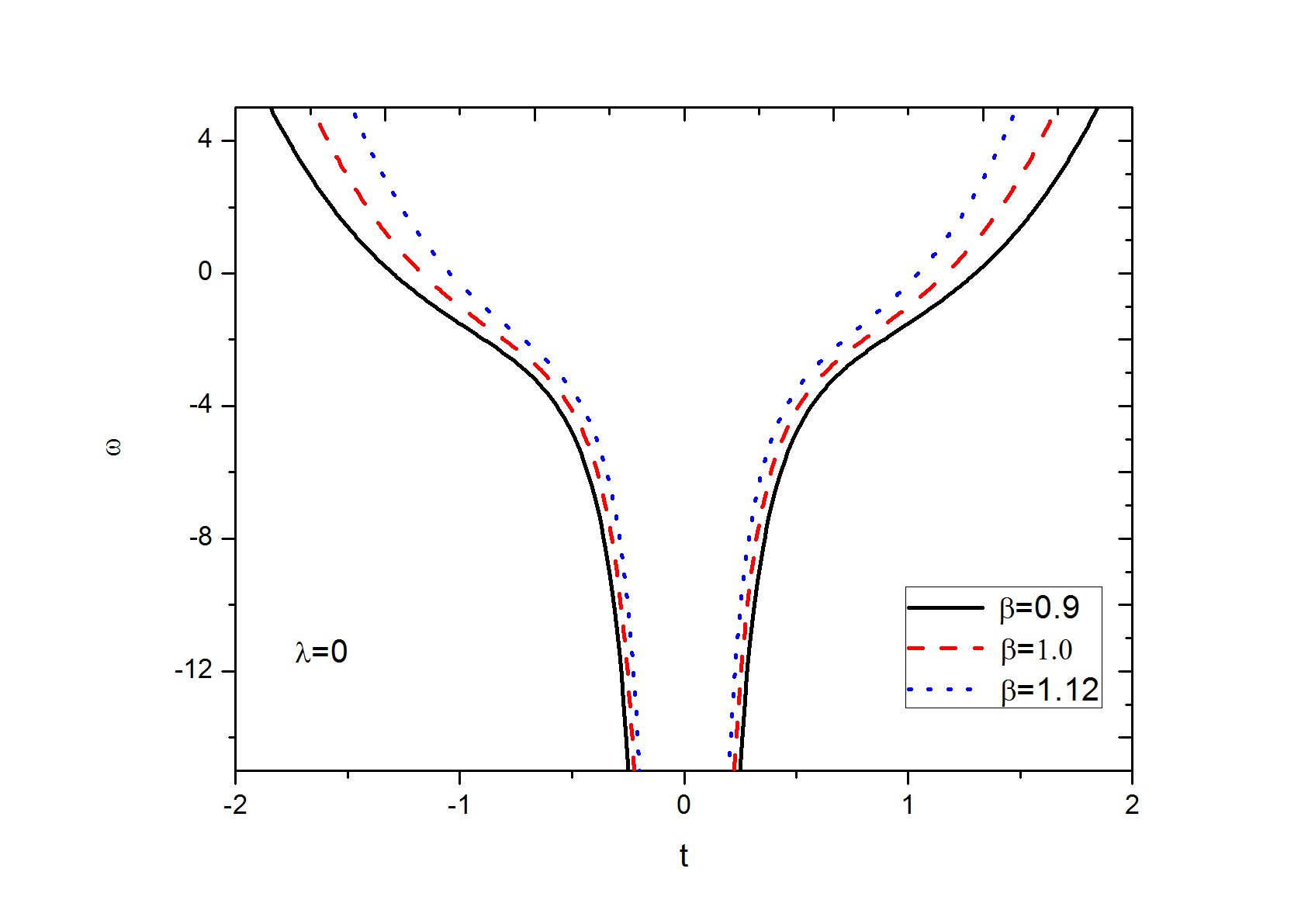}
\caption{EoS parameter for four representative values of the parameter $\alpha$ for $k=0.5$. } 
\endminipage 
\minipage{0.40\textwidth}
\includegraphics[width=65mm]{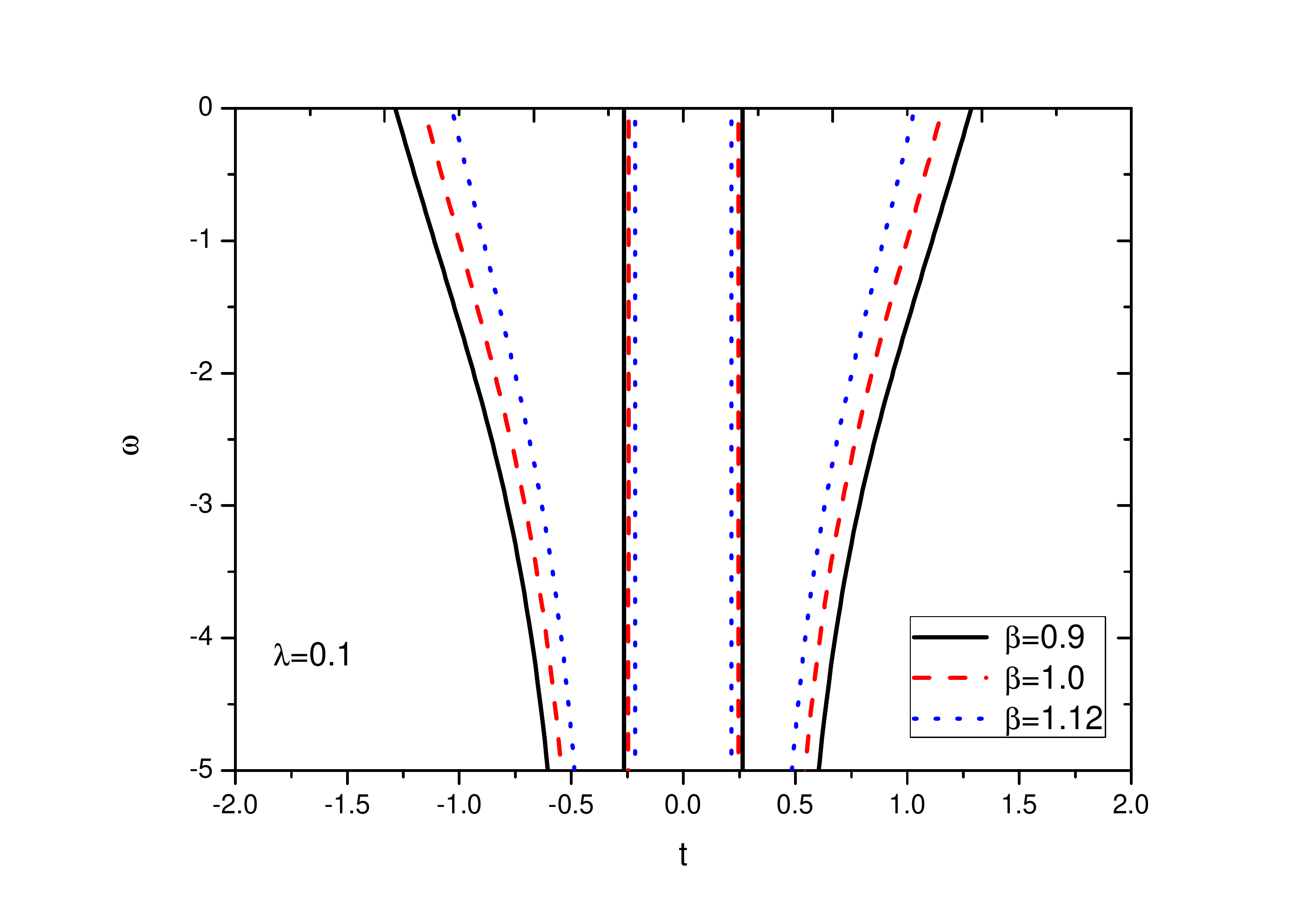}
\caption{EoS parameter for four representative values of the parameter $\alpha$ for $k=1$. }
\endminipage\\
\minipage{0.40\textwidth}
\centering
\includegraphics[width=65mm]{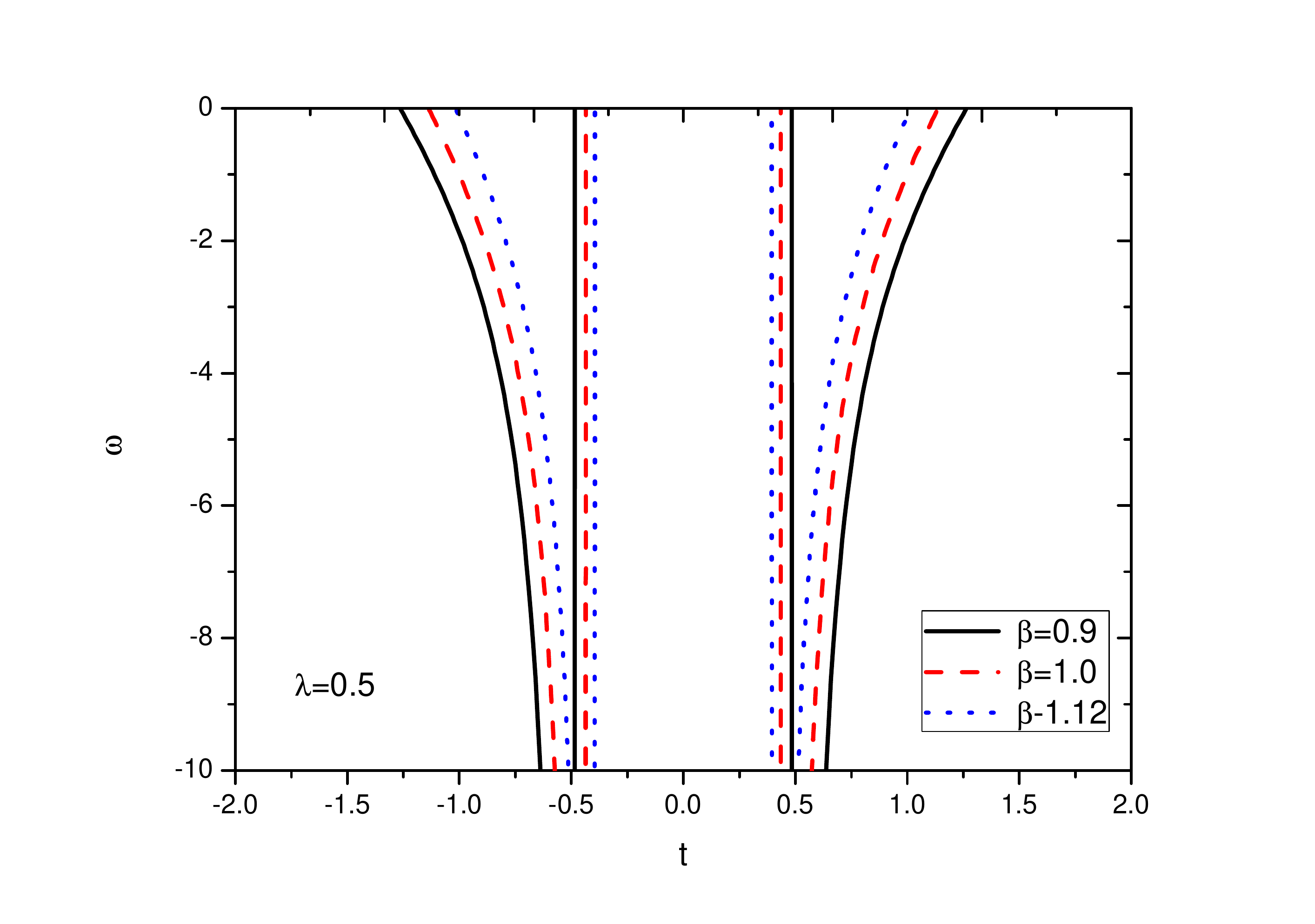}
\caption{EoS parameter for four representative values of the parameter $\alpha$ for $k=5$. } 
\endminipage 
\minipage{0.40\textwidth}
\includegraphics[width=65mm]{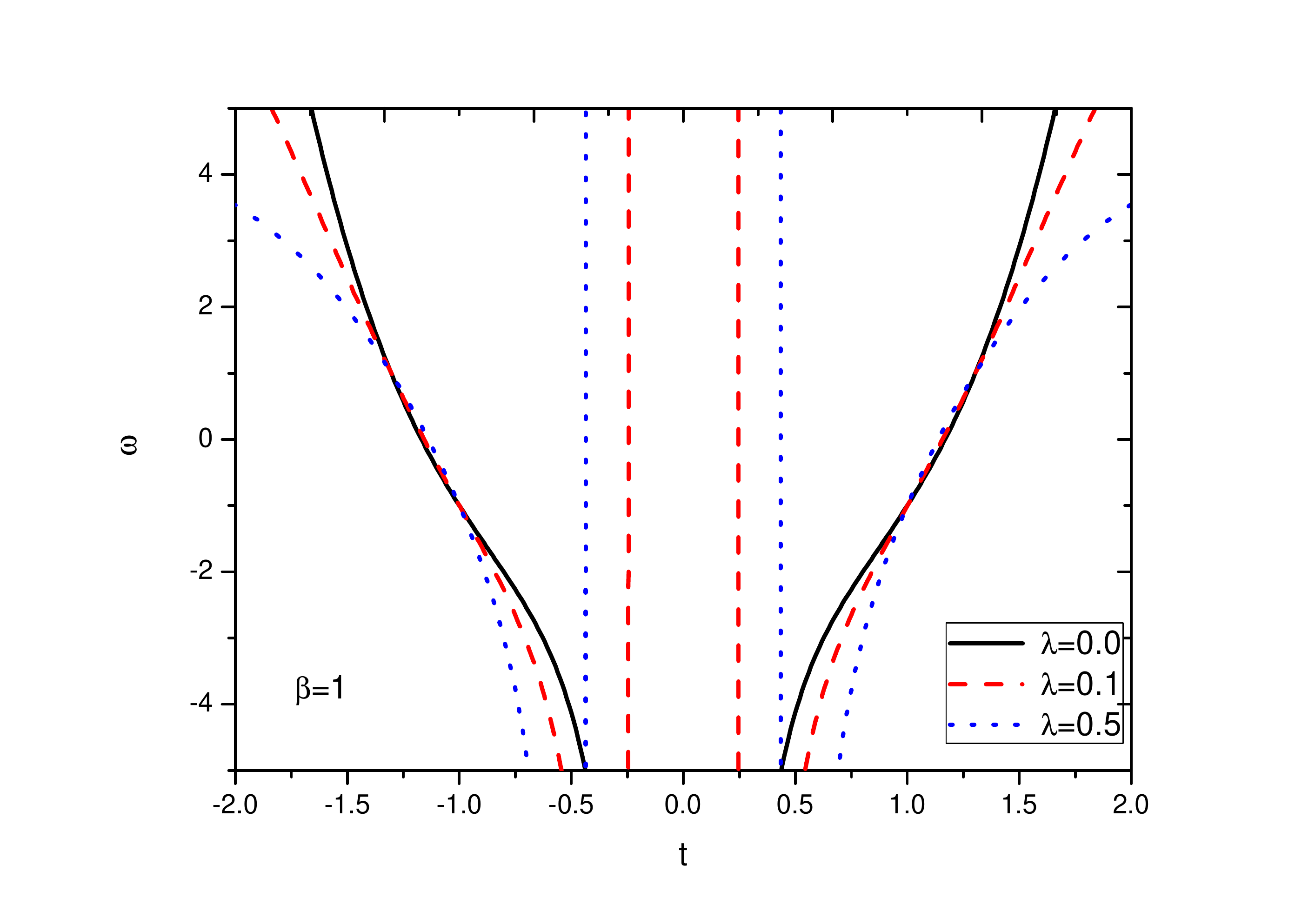}
\caption{EoS parameter for four representative values of the parameter $\alpha$ for $k=9$. }
\endminipage
\end{figure}



The null energy condition in this model can be calculated as
\begin{equation}
\rho+p=-\frac{2\beta^2(1+4\lambda)\left(1-\beta^2t^2\right)}{(1 + \beta^{2}t^{2})^{2}[{( 1 + 3\lambda)^{2} - \lambda^{2}}]}.
\end{equation}
This is a negative quantity and signifies a violation of the null energy condition.
\section{Conclusion}
We have investigated some bouncing models in the framework of an extended theory of gravity. In the gravitational action of the extended gravity theory the usual Ricci scalar $R$ of GR action is replaced by  a sum of $R$ and a term proportional to the trace $T$ of the energy momentum tensor. The extra term in the action provides an anti-gravity effect and may be able to explain the late time cosmic acceleration. The non singular bouncing models investigated in this work also favour late time cosmic speed up phenomenon besides being able to provide a viable bouncing scenario at $t=0$. We have studied the dynamical evolution of the equation of state parameter in the modified theory of gravity keeping in view the bouncing scenario. The dynamics of the models are greatly affected by the parameters of the model. While an increase in the coupling parameter of the extended gravity theory decreases the rate of dynamical evolution, an increase in the bouncing scale factor parameter raises the rate of dynamics. The behaviour near the bounce is mostly decided by the coupling parameter of the extended gravity theory and is least affected by the bounce scale factor.

\section*{Acknowledgement}
SKT thanks IUCAA, Pune (India) for providing support during an academic visit where a part of this work is carried out.

\end{document}